\documentclass[]{JHEP} 
\def\beq{\begin{equation}}
\def\eeq{\end{equation}}
\def\bar{\begin{eqnarray}}
\def\ear{\end{eqnarray}}
\def\id{I\kern-4.5pt I}
\def\q{$Q$}

\usepackage{epsfig,amssymb}
\newcommand\fverb{\setbox\pippobox=\hbox\bgroup\verb}
\newcommand\fverbdo{\egroup\medskip\noindent%
            \fbox{\unhbox\pippobox}\ }
\newcommand\fverbit{\egroup\item[\fbox{\unhbox\pippobox}]}
\newbox\pippobox
\title{From Q-Walls to Q-Balls }

\author{R. B. MacKenzie\\
    Groupe de physique des particules, Laboratoire Ren\'e-J.-A.-L\'evesque, Universit\'e de Montr\'eal, C. P. 6128, succ. centre-ville, Montr\'eal, Qu\'ebec, Canada, H3C 3J7 \\
    E-mail: \email{rbmack@lps.umontreal.ca}}
    
\author{M. B. Paranjape\thanks{permanent address: Groupe de physique des particules, Laboratoire Ren\'e-J.-A.-L\'evesque, Universit\'e de Montr\'eal, C. P. 6128, succ. centre-ville, Montr\'eal, Qu\'ebec, Canada, H3C 3J7 }\\
    Departamento de F{\'{\i}}sica Te{\'o}rica, Facultad de Ciencias, Universidad de Zaragoza, 50009, Zaragoza, Espa\~na\\
    E-mail: \email{paranj@lps.umontreal.ca}}

\received{\today}       
\accepted{\today}       

\preprint{\hepth{*******}}  

\abstract{
We study $Q$-ball type solitons in arbitrary spatial dimensions in the
setting recently described by Kusenko, where the scalar field
potential has a flat direction which rises much slower than $\phi^2$.
We find that the general formula for energy as a function of the
charge is, $E_d\sim Q_d^{(d/d+1)}$ in spatial dimension $d$.  We show
that the Hamiltonian governing the stability analysis of certain
$Q$-wall configurations, which are one dimensional $Q$-ball solutions
extended to planar, wall-like configurations in three dimensions, can
be related to supersymmetric quantum mechanics. $Q$-wall and
$Q$-string (the corresponding extensions of 2 dimensional $Q$-balls in
3 spatial dimensions) configurations are seen to be unstable, and
will tend to bead and form planar or linear arrays of 3 dimensional
$Q$-balls.  The lifetime of these wall-like and string-like
configurations is, however, arbitrarily large and hence they could be
of relevance to cosmological density fluctuations and structure
formation in the early Universe. }

\dedicated{Dedicated to our loves, our lives\ldots\\our wives.}
\begin{document} 

\maketitle 

\section{Introduction}

Non-topological solitons were first described in the mid-70's \cite{fls} in the context of scalar field models with two fields, symmetry breaking and a conserved charge.  They reappeared in a somewhat different guise in the mid 80's in theories without spontaneous symmetry breaking as considered by Coleman \cite{c} and the name $Q$-balls was invented.  In the mid 90's they have re-surfaced in connection with models of scalar fields where the potential has flat or quasi-flat directions, which is generic in models of supersymmetry breaking, as described by Kusenko \cite{k}.  

The original non-topological solitons and {\q}-balls  had an energy which behaved as 
\beq
E\sim \mu Q
\eeq
for large conserved charge $Q$.  If $\mu $ is less than the mass of a perturbative, charged excitation of unit charge, then the {\q}-balls are clearly stable against decay into perturbative charged particles.  Indeed, Coleman \cite{c} defined {{\q}} balls as the minimal energy classical configurations of fixed charge {{\q}} and was able to show their absolute, classical stability.  However, it was soon realized that linear terms induced via quantum corrections in the effective potential would lead to the decay of these {\q}-balls \cite{ccgm}.  

The recently-described {\q}-balls do not suffer this fate.  They have
an intriguing energy-charge dispersion, $E\sim Q^{(3/4)}$ for large
{\q} in 3 spatial dimensions.  Hence these {\q}-balls are arbitrarily
tightly bound compared to the energy of an aggregate of perturbative
quanta of equal charge. We note that such an asymptotic energy-charge
dispersion with the non-integer power $3/4$ also arises in another
context.  For solitons in models with non-trivial topological Hopf
windings \cite{nf}, there is an energy bound, which is not attained,
but behaves as $Q^{(3/4)}$.  It would be interesting to find a model
where a ``Hopf soliton", perhaps of the {\q}-ball, type can exactly
saturate such a bound.  Indeed, we will find below exactly solvable
{\q}-balls which do saturate a similar bound in 1 spatial dimension.

In this paper we investigate extended objects of the {\q}-ball
type. These are the solitons of one- and two-dimensional models
embedded in three dimensions, forming {\q}-walls and {\q}-strings,
respectively.  Similar structures have been studied recently in \cite{bs,pv} both
analytically and using numerical simulations.  The cases discussed concern 
{\q}-balls which have an energy-charge dispersion that grows linearly
with the charge, $E\sim\mu Q$.  It is found that collisions of such
\q-balls can give rise to \q-string like states.  In \cite{pv} stable
\q-ring solitons are described.  Here stability is imbued via an additional
topological twisting of the phase as one goes around the ring.  We find our
study is complementary to these studies.  

In the next section, we study {\q}-solitons in arbitrary spatial
dimension $d$. The energetics of these objects are interesting: the
energy $E_d$ of a soliton in $d$ dimensions depends on its charge
$Q_d$ as $E_d\sim Q_d^{d/(d+1)}$. The lower the dimension, the more
energetically favourable the {\q}-soliton over a conventional
configuration of charge $Q_d$. At first glance, this would seem to
indicate that extended objects in three dimension would be
energetically preferred over {\q}-balls. However, as we shall see,
this turns out not to be the case.

In Section 3, we consider embedded lower-dimensional {\q}-solitons in
three dimensions. The energy $E_2$ and charge $Q_2$
of a two-dimensional soliton,
for instance, are the energy and charge per unit length of the
corresponding {\q}-string. The energy of a {\q}-string of length $L$
and of fixed three-dimensional charge varies as $L^{1/3}$. Thus the
tension varies as $L^{-2/3}$, reducing rapidly with length. However,
the energy per unit length also varies as $L^{-2/3}$, so the natural
time scale for oscillations of {\q}-strings is equivalent to that of
topological strings (whose energy per unit length and tension are
independent of $L$).

In Section 4 we study the stability of {\q}-strings or walls against
``beading'', {\it i.e.,} the formation along the length of the string
(or surface of the wall) of clumps of higher charge (and
energy) per unit length (or surface) which would ultimately result in
the formation of a one-dimensional chain (or two-dimensional sheet)
of non-interacting balls.

We study the stability of walls, in particular, since there the
analysis is much simpler, essentially because the one-dimensional
soliton can be found analytically for certain potentials. In studying
small oscillations about such a solution, the eigenvalue equation
determining the frequency of these oscillations can be related to a
Schr\"odinger equation in supersymmetric quantum mechanics. Applying
known results from that theatre enables us to identify a single
unstable mode for the fluctuation, two zero-modes (one of which
corresponds to translation of the wall) and a spectrum of stable
modes. The unstable mode can have an arbitrarily large characteristic
time, so the wall can be extremely long-lived. We expect the same
behaviour for strings, but have not yet done the analysis in that case.

\section{{\q}-balls in any dimension}
Consider a charged scalar field in $d$ spatial dimensions, the Lagrangian governing its behaviour is
\beq
{\cal L}=\int d^dx\left( {1\over 2}\partial_\mu\phi_a
  \partial^\mu\phi_a -V(|\phi |)\right)
\eeq
where $a=1,2$, $|\phi |=\sqrt{\phi_1^2+\phi_2^2}$ and we use a simple $SO(2)$ global gauge invariance whose conserved charge corresponds to {\q}.  The conserved charge is normalized as
\beq
Q=\int d^dx\,\epsilon_{ab}\phi_a\partial_t\phi_b.
\eeq
The scalar potential, inspired by supersymmetric models, satisfies
\beq
V(|\phi |)=\cases{{1\over 2}m^2 |\phi |^2&$|\phi |\rightarrow 0$\cr \Lambda &$|\phi |\rightarrow\infty$}
\eeq
where $\Lambda$ is a constant which is determined by the scale of supersymmetry breaking.  The minimum of the potential occurs at $|\phi |=0$ and the charge symmetry is unbroken.  {\q}-ball solutions are found with the ansatz
\beq
\phi_1 +i\phi_2 = e^{i\omega t}\varphi(r)
\eeq
where $\omega$ is a constant, $r$ is the radial coordinate and $\varphi (r)$ is real.  We immediately obtain
\beq
Q=\omega\int d^dx\, \varphi^2 (r).
\eeq
The profile taken for $\varphi (r)$ satisfies the asymptotic conditions
\beq
\varphi (r) =\cases{\phi_0 & $r\approx 0$\cr 0&$r\rightarrow\infty$ }
\eeq
where $\phi_0$ is taken large enough so that the asymptotic behaviour of the potential is valid.  Hence the {\q}-ball interpolates from large non-vacuum values around $\phi_0$ in its center to $\varphi =0$, the perturbative vacuum, at infinity.  The existence of stable solutions to the equations of motion for $\varphi (r)$ in 3 spatial dimensions is the content of the previous papers \cite{c,c2} and is easiest seen by the classic continuity arguments of Coleman \cite{c}.  Here we re-do the analysis in arbitrary spatial dimensions $d$.  

The equations of motion for $\varphi (r)$ are
\beq
\nabla^2\varphi (r) -V^\prime (\varphi (r))+\omega^2\varphi (r)=0\label{eqom}
\eeq
which, given the form of the potential, become
\bar
\nabla^2\varphi (r) +\omega^2\varphi (r)&=&0\quad r\approx 0\cr
\nabla^2\varphi (r)-m^2\varphi (r)&=&0\quad r\approx\infty 
\ear
using the assumed asymptotics of the potential.  The solutions of
these equations are immediate, but dimension dependent.  For small $r$
we get spherical Bessel functions of order 0 in 3 dimensions, ordinary
Bessel functions of order 0 in 2 dimensions, and trigonometric
functions in 1 dimension, with argument $\omega r$.  For large $r$ the
argument simply changes to $imr$, which is imaginary.
\subsection{$d=3$}
The asymptotic solutions are
\beq
\varphi (r)=\cases{\phi_0j_0 (\omega r)&$r\approx 0$\cr \varphi_0k_0 (m r)&$r\approx \infty $}\label{as}
\eeq
where $\phi_0$ is a constant that is fixed by the total charge of the solution, $\varphi_0$ is a dependent constant and $j_0(\omega r)$ and $k_0(m r)$ are the spherical Bessel function of order 0 of real and imaginary argument respectively.
The asymptotic solutions interpolate from one to the other in a relatively small region, and it is difficult to obtain analytic solutions, except in the 1 dimensional case.  However, the continuity argument of Coleman easily establishes the existence of a solution.  

The idea is to first observe that the equation of motion \ref{eqom} can be interpreted to describe the classical motion of a ``particle" in one dimension of position  $\varphi$, in the potential $-V(\varphi ) +{1\over 2}\omega^2\varphi^2$ and in the presence of a specific frictional term that is time dependent coming from the Laplacian, and where $r$ plays the role of time. Indeed the equation of motion is equivalently written as
\beq
\ddot \varphi (r) -V^\prime (\varphi (r)) +\omega^2 \varphi (r)=-{2\over r}\dot\varphi (r)
\eeq
where $\dot\varphi (r) =\partial_r\varphi (r)$.   The potential is negative for small $\varphi$, and is approximately a downward opening parabola of curvature $-m^2$, whose maximum passes through the origin at $\varphi (r) =0$.  It turns over, after achieving a minimum (negative) value, into an upward opening parabola of curvature $\omega^2$ for large enough values of  $\varphi (r)$. Clearly, eventually the potential becomes positive.   Now if we start with an initial value $\phi_0$ for which $V(\phi_0)$ is negative, the ``particle" will never rise up high enough to reach the origin, even in infinite time, and will perform damped oscillations till it settles down to the global minimum value of $-V(\varphi ) +{1\over 2}\omega^2\varphi^2$.  On the other hand if we start with a large initial value of $\phi_0$, then since the ``motion" is governed arbitrarily well by the spherical Bessel function $j_0(\omega r)$, the ``particle" will attain a given value close to the point where the full potential $-V(\varphi ) +{1\over 2}\omega^2\varphi^2$ vanishes, in a fixed time, independent of the initial value $\phi_0$ which is of the order of $1/\omega$.  The velocity at this point will be proportional to $\phi_0$.  Hence the subsequent evolution can be arranged to commence with an arbitrarily high initial velocity by choosing $\phi_0$ larger and larger.  It is clear then, that there exists an initial value for $\phi_0$ so that the ``particle"will arrive at this point with sufficient velocity that the subsequent motion will rise up and go over the maximum at $\varphi=0$ in finite time.  Hence, by continuity, there exists an initial value for $\phi_0$ which will exactly achieve the maximum as time goes to infinity, {\it i.e.,} for $r\rightarrow\infty$.  Such a configuration is the solution and it will give the appropriate interpolation between the two asymptotic solutions \ref{as}.  Clearly the value of $\varphi_0$ will be governed by the value of $\phi_0$ which in turn will be fixed by the value $\omega$. Hence the total charge {\q} can be thought of as a function of $\omega$ alone, or conversely all of the parameters depend only on the total charge \q.  

The energy of such a configuration can be approximated as
\bar
E&\approx &\int_{r<1/\omega} \hskip-.7cm{d^3x\,  \big( \omega^2\varphi^2(r) +|\nabla\varphi (r)|^2 +\Lambda\big)}\cr
&=&\int_{r<1/\omega} \hskip-.7cm{d^3x\,  \big( 2\omega^2\varphi^2(r) +\Lambda\big)}= {\alpha \phi_0^2\over\omega} +{\beta\Lambda\over\omega^3}
\ear
where we neglect the contributions from the surface and the exterior of the {\q}-ball, and where $\alpha$ and $\beta$ are positive constants.  This expression is of course reasonable only for {\q}-balls with large charge {\q}.  
Using 
\beq
Q=\omega\int d^3\, x \varphi^2(r)\approx \omega \int_{r<1/\omega} \hskip-.7cm{d^3 x\,\phi_0^2j_0^2(\omega r)}={\gamma\phi_0^2\over \omega^2}
\eeq
where $\gamma$ is another positive constant, we get
\beq
E\approx {\alpha Q\omega\over \gamma} + {\Lambda\beta\over\omega^3}
\eeq
Imposing $dE/d\omega =0$ yields
\beq
{\alpha Q\over\gamma}={3\Lambda\beta\over\omega^4}
\eeq
which gives
\bar
\omega &\sim & Q^{-1/4}\cr
\phi_0 &\sim & Q^{1/4}\cr
 E&\sim & Q^{3/4} .
\ear
\subsection{$d=2$}
Here the solution for small $r$ is $\phi_0J_0(\omega r)$ where $J_0(\omega r)$ is the ordinary Bessel function of order 0, and for large $r$ the solution is $\varphi_0K_0(mr)$, where $K_0(mr)$ is the Bessel function of order zero of imaginary argument.  The energy is then approximately given by
\bar
E_2&\approx &\int_{r<1/\omega} \hskip-.7cm{d^2x\,  \big( \omega^2\varphi^2(r) +|\nabla\varphi (r)|^2 +\Lambda\big)}\cr
&=& {\alpha Q_2\omega\over\gamma} +{\beta\Lambda\over\omega^2}
\ear
which yields upon imposing $dE_2/d\omega =0$
\bar
\omega &\sim & Q_2^{-1/3}\cr
\phi_0 &\sim & Q_2^{1/3}\cr
 E_2&\sim & Q_2^{2/3} .
\ear
Here $E_2$ and $Q_2$ are the appropriate 2 dimensional energy and charge respectively, which in terms of 3 dimensional energy and charge are the energy per unit length and the charge per unit length. 

\subsection{$d=1$}\label{2.3}
Here the equation of motion can be reduced to quadratures, and it can be integrated exactly in many cases.  It is more convenient to use the normal spatial coordinate $x\in [-\infty ,\infty ]$ rather than the radial coordinate $r$.  The equation of motion for $\varphi (x)$ where $\phi_1(x)+i\phi_2(x) =e^{i\omega t}\varphi (x)$ is
\beq
{d^2\varphi (x)\over dx^2} -V^\prime (\varphi (x)) +\omega^2\varphi (x) = 0
\eeq
which yields
\beq
\int_{\phi_0}^{\varphi (x)}{d\varphi\over\pm\sqrt{V(\varphi ) -\omega^2\varphi^2}} =\int_{x_0}^{x}dx.\label{quadrature}
\eeq
Here $x_0$ gives the position of the {\q}-ball and the $\pm$ is chosen depending on whether $x> x_0$ or $x< x_0$.  The square root should not become imaginary thus we must have that $V(\varphi ) >\omega^2\varphi^2$.  Since for large values of $\varphi$ the potential becomes flat and the required inequality is violated, there is a fixed maximum value for $\phi_0$ given by the condition 
$V(\phi_0 ) =\omega^2\phi_0^2$.  There is no ``friction" term coming from the Lapalcian in 1 dimension, hence conservation of energy dictates that this value of $\phi_0$ is the starting point for $\varphi_0(x)$.   Small $\omega$ implies large values of $\phi_0$ which gives approximately that 
$\phi_0^2\approx{\Lambda /\omega^2}$ that is $\phi_0\sim 1/\omega$.  The quadrature \ref{quadrature} is not particularly useful in the case of a general potential $V(\varphi (x))$.  The solution for $x\approx x_0$ is $\varphi (x)=\phi_0\cos ( \omega (x-x_0))$ while for $|x-x_0|>>\pi /2\omega$ is $\varphi (\pm x)=\varphi_0 e^{\mp mx}$.  The charge is given by
\beq
Q_1=\omega\int dx\varphi^2 (x)\approx \omega\phi_0^2\int_{x_0-(\pi /2\omega )}^{x_0+(\pi /2\omega )}\hskip-1.2cm{dx\, \cos^2(\omega (x-x_0)) } ={\phi_0^2\pi\over 2}
\eeq
while the energy is
\bar
E_1&=&\int dx \big( \omega^2\varphi^2 (x) +(\partial_x\varphi (x))^2 + V (\varphi (x) )\big) \cr
&=&\int dx\big( 2\omega^2\varphi^2 +V(\varphi (x) ) -\varphi (x) V^\prime (\varphi (x) )\big)\cr
&\approx & \int_{|x-x_0|<(\pi /2\omega )} {\hskip-1.8truecm dx\big( 2\omega^2\varphi^2 +\Lambda \big)}\cr
&=&2\omega^2\phi_0^2\Big( \int_{|x-x_0|<(\pi /2\omega )}{\hskip-1.8truecm dx\cos^2(\omega (x-x_0))}\Big) +{\Lambda\pi\over\omega}\cr
&=&{2\pi\omega\phi_0^2\over 2}+{\Lambda\pi\over\omega }.
\ear
Imposing 
\beq
{dE_1\over d\omega}=0
\eeq
yields
\beq
\phi_0^2 -{\Lambda\over\omega^2}={2Q_1\over\pi}-{\Lambda\over\omega^2}=0
\eeq
implying
\beq
\omega^2={\pi\Lambda\over 2 Q_1}.
\eeq
This gives,
\bar
\omega &\sim & Q_1^{-1/2}\cr
\phi_0 &\sim & Q_1^{1/2}\cr
 E_2&\sim & Q_1^{1/2} .
\ear
\subsection{Arbitrary spatial dimension}
Here we record the corresponding formulae in arbitrary spatial
dimension $d$.  The $Q_d$-ball solution will be to the appropriate
generalization of the free spherical wave in its centre with argument
$\omega r$ while from its edge it will evolve into the exponentially
decaying type solution of mass $m$ outside, {\it i.e.,} the same function but of imaginary argument $imr$.  One easily finds
\bar
Q_d&\approx &{\gamma_d\omega\phi_0^2\over\omega^d}= {\gamma_d\phi_0^2\over\omega^{(d-1)}}\cr
E_d&\approx &\alpha_d \phi_0^2\omega^{(2-d)}+{\beta_d\Lambda\over\omega^d}\cr
&=&{\alpha_d Q_d\omega^{(d-1)}\over\gamma_d} \omega^{(2-d)}+{\beta_d\Lambda\over\omega^d}\cr
&=&{\alpha_d Q_d\omega\over\gamma_d} +{\beta_d\Lambda\over\omega^d}
\ear
where $\alpha_d , \beta_d$ and $\gamma_d$ are the corresponding constants in $d$ dimensions.
Imposing $dE_d/d\omega =0$ yields in a straightforward manner
\bar
\omega &\sim & Q_d^{-1/(d+1)}\cr
\phi_0 &\sim & Q_d^{1/(d+1)}\cr
 E_d&\sim & Q_d^{d/(d+1)} .
\ear
\section{Energetics of {\q}-strings and {\q}-walls}
Suppose initial conditions have created a large {\q}-wall or {\q}-string in 3 dimensions.  We will observe here that such structures have interesting energetics.  We address the question of stability of such objects in the next section.  
\subsection{{\q}-strings}
We define {\q}-strings as linear configurations in 3 dimensions whose cross section corresponds to a 2 dimensional $Q_2$-ball, with a fixed charge per unit length $Q_2$.  Such a configuration will simply terminate at two end points or close on itself yielding a closed {\q}-string configuration.  We will consider for convenience the second situation, of a closed {\q}-string of length $L$.  We will assume that $L$ is sufficiently long so that gradient energies due to the gradual bending that is necessary to close the {\q}-string are negligible. The relevant conserved charge of a closed {\q}-string then is given by
\beq
Q=Q_2L
\eeq
while the energy is given by
\beq
E=E_2L.
\eeq
Replacing for $E_2\sim Q_2^{2/3}=(Q/L)^{2/3}$ we get
\beq
E\sim Q_2^{2/3}L=Q^{2/3}L^{1/3}.
\eeq
Hence a {\q}-string of total charge {\q} satisfies an energy-charge dispersion relation  
\beq
E_{Q{\rm -string}}\sim Q^{2/3}.
\eeq
This does not mean that the {\q}-string is even more tightly bound than a {\q}-ball with the same charge, which satisfies $E\sim Q^{3/4}$.
Overall scaling of the {\q}-string always lowers its energy.  The energy-length dispersion is
\beq
E_{Q{\rm -string}}\sim L^{1/3}
\eeq
so that as the length decreases, so does the energy.  Indeed the {\q}-string will shrink and reduce its energy until its length is comparable to its width, at which point the description in terms of a {\q}-string is no longer valid.  

The energy-length dispersion is however, remarkable.  The restoring force which tends to cause the {\q}-string to shrink
\beq
F_{Q{\rm -string}}=-{dE\over dL}\sim -{1\over 3}{Q^{2/3}\over L^{2/3}}
\eeq
disappears as $L$ gets very large.  This is in contradistinction to the case of a topological string, such as a cosmic string or vortex loop, where the 
energy-length dispersion is $E_{\rm top.-string}=\mu L$, where $\mu$ is the mass per unit length, giving a constant restoring force.  Nevertheless, the time scale for shrinking of a {\q}-string is not markedly different, for a large {\q}-string the mass per unit length also decreases accordingly with its length, compensating for the reduced restoring force.  

The time scale for shrinking of a topological string is governed by the effective Lagrangian
\beq
{\cal L}_{\rm top.-string}={1\over 2}\mu L{\dot L^2\over 4\pi^2}-\mu L =\mu \big( {L\dot L^2\over 8\pi^2}-L\big)
\eeq
while for a {\q}-string we have
\beq
{\cal L}_{Q\rm -string}={1\over 2}\big( {\alpha Q_2^{2/3}\over L^{2/3}}\big) L{\dot L^2\over 4\pi^2}-\big( {\alpha Q_2^{2/3}\over L^{2/3}}\big) L =\alpha Q_2^{2/3} \big( {L^{1/3}\dot L^2\over 8\pi^2}-L^{1/3}\big) .
\eeq
We see that neither $\mu$ nor $\alpha Q_2^{2/3}$ play a role in the dynamics.  Conservation of energy for the topological string gives
\beq
{L\dot L^2\over 8\pi^2}+L=L_0
\eeq
while for the {\q}-string
\beq
{L^{1/3}\dot L^2\over 8\pi^2}-L^{1/3}=L_0^{1/3},
\eeq
where $L_0$ is the initial length in each case.  With the change of variables $L=v^2$ for the topological string and $L=u^6$ for the {\q}-string the equations become easily integrable, yielding the time for contraction $T_{\rm top.-string}=L_0/4\sqrt 2$ and $T_{Q\rm -string}=45L_0/32\sqrt 2$.

We will see in the next section that the {\q}-string and the {\q}-wall are both unstable to beading into arrays of smaller \q-balls.  If the {\q}-string were to spontaneously decay into a linear array of $N$, {\q}-balls each of charge $q$, we have
\beq
Q=Nq
\eeq
from conservation of charge, while the energy of such a final state is
\beq
E=\beta q^{3/4}N=\beta Q^{3/4}N^{1/4}.
\eeq
Evidently such a state has higher energy than the absolute minimum of energy in the sector of charge {\q}, given by the {\q}-ball of charge {\q}, ($N=1$), which has energy $E=\beta Q^{3/4}$.  But it will be an intermediate state with less energy than the {\q}-string if (adopting the notation of Section 2.4 to avoid confusion)
\beq
\beta_3Q^{3/4}N^{1/4}<\beta_2 Q^{2/3}L^{1/3}.
\eeq
This implies
\beq
N<\big({\beta_2\over\beta_3}\big)^4{L^{4/3}\over Q^{1/3}}
\eeq
which gives $N/L<(\beta_2/\beta_3 )^4Q_2^{-1/3}$.  As the size of each
smaller {\q}-ball is approximately $R\sim L/N$ thus
$R>(\beta_3/\beta_2 )^4Q_2^{1/3}$, {\it i.e.,} the {\q}-string will be stable against decay into a chain of $N$, {\q}-balls if the size of each {\q}-ball is smaller than $Q_2^{1/3}\sim 1/\omega$, which is just the width of the original {\q}-string.
\subsection{{\q}-walls}
We can repeat the above analysis for the case of a closed {\q}-wall of total 3 dimensional charge {\q} and with a surface area $L^2$.  Then 
\beq
Q=L^2Q_1
\eeq
and 
\beq
E_{Q\rm -wall}=L^2E_1 =L^2\beta_1Q_1^{1/2}=\beta_1L^2 (Q/L^2)^{(1/2)}=\beta_1 Q^{(1/2)}L.
\eeq
This is to be compared with the energy of a topological domain wall,
\beq
E_{\rm top.-wall}=\mu L^2
\eeq
where $\mu$ is the mass per unit area.  For a spherical surface the effective Lagrangian describing the dynamics of a spherical topological domain wall is 
\beq
{\cal L}_{\rm top.-wall}={1\over 2}\mu L^2{\dot L^2\over 4\pi}-\mu L^2
\eeq
while for a spherical {\q}-wall we get
\beq
{\cal L}_{Q\rm -wall}={1\over 2}\beta_1Q^{1/2}L{\dot L^2\over 4\pi} -\beta_1Q^{1/2}L.
\eeq
These yield the conservation law for the topological domain wall 
\beq
{1\over 8\pi}L^2\dot L^2 +L^2=L_0^2
\eeq 
while for the {\q}-wall
\beq
{1\over 8\pi}L\dot L^2 +L=L_0,
\eeq 
where $L_0$ is the initial value of $L$.  These are easily integrable and yield  the time for contraction to be
\beq
T_{\rm top.-wall}={L_0\over 2\sqrt{2\pi}}
\eeq
and 
\beq
T_{Q\rm -wall}=\sqrt{\pi\over 2}{L_0\over 4}
\eeq
for the topological domain wall and for the {\q}-wall respectively.  Again we see that the time for contraction is about the same in the two cases.  

If the {\q}-wall spontaneously beads up into a locally planar array of $N$ smaller {\q}-balls of charge $q$, we have 
\beq
Q=Nq
\eeq
and 
\beq
E=N\beta_3 q^{3/4}=\beta_3Q^{3/4}N^{1/4}.
\eeq
Then energy considerations dictate
\beq
\beta_3N^{1/4}Q^{3/4}<\beta_1Q^{1/2}L,
\eeq
{\it i.e.,}
\beq
\big({\beta_3\over\beta_1}\big)^4Q_1^{1/2}<\big({L^2\over N}\big)^{1/2}
\eeq
which says the size of each small {\q}-ball must be greater than the width of the {\q}-wall.
\section{Stability}
The general rate of decay is governed by the stability analysis of the fluctuations about the {\q}-wall or {\q}-string configuration.  The total Lagrangian is expanded to second order in the fluctuations about the soliton configuration. The first order terms vanish because the soliton is a solution of the full 3 dimensional equations of motion.  The second order truncation yields a generalized normal mode problem.  Real frequencies lead to oscillatory behaviour while imaginary frequencies yield exponential growth or decay.  In the latter case, the lifetime is taken to the inverse of the magnitude of the imaginary frequency, which is the time scale for the instability leading to beading.
We will analyze the problem of the stability of a {\q}-wall for a specific potential which leads to an analytically accessible problem.  
\subsection{Analytical {\q}-walls}\label{aqw}
As we have seen in section \ref{2.3}, the expression for a {\q}-ball in 1 dimension can be brought to quadrature.  For many choices of the potential the resulting integral can be computed analytically \cite{khare}.  The stability analysis of the corresponding {\q}-wall is related to the elegant theory of exactly solvable quantum mechanical systems, supersymmetric quantum mechanics and shape invariant potentials \cite{cks}.  
Consider the potential
\beq
V(\varphi )=\omega^2\big( {5\over 2}\varphi^2 -2{\varphi^3\over\phi_0}\big) .\label{potential}
\eeq
Then 
\beq
-V(\varphi )+{1\over 2}\omega^2\varphi^2=0
\eeq
implies
\beq
\varphi^2=0\quad{\rm or}\quad\varphi =\phi_0.
\eeq
The non-trivial zero is fixed at $\phi_0$ and the {\q}-ball solution will start from this value for $\varphi$ at its centre.  The equation of motion is
\beq
-\varphi^{\prime\prime}_0(x)-\omega^2\varphi_0(x) +V^\prime (\varphi_0(x))=0
\eeq
which becomes
\beq
-\varphi^{\prime\prime}_0(x) +\omega^2\big(4\varphi_0(x)-6{\varphi_0^2(x)\over\phi_0}\big)=0.\label{varphi0}
\eeq
The quadrature \ref{quadrature} is integrable in this case, with solution
\beq
\varphi_0(x)=\phi_0{\rm sech}^2 (\omega x).
\label{solution}
\eeq
It is remarkable that the stability problem around such a {\q}-wall is related to the corresponding suspersymmetric quantum mechanical system, as we will now show.  
\subsection{Second-order Lagrangian}
The stability analysis proceeds with the second-order expansion of the
total Lagrangian about the {\q}-wall configuration.  We will consider
an infinite planar {\q}-wall, which then gives that \ref{solution} is
an exact solution of the full 3 dimensional equations.  This will be
an arbitrarily close approximation for a large but finite closed
{\q}-wall surface.  The second-order Lagrangian for 
$\phi (\vec x,t)=e^{i\omega t}\varphi (\vec x,t)=e^{i\omega
  t}(\varphi_0(x)+\varphi_1 (\vec x,t) +i\varphi_2(\vec x,t))$ is
\beq
{\cal L}^{[2]}=\int d^3 x {1\over 2}(\partial_t\varphi_a-\omega\epsilon_{ab}\varphi_b)
(\partial_t\varphi_a-\omega\epsilon_{ab}\varphi_b)+ {1\over 2}\partial_i\varphi_a\partial_i\varphi_a -{1\over 2}V^{\prime\prime}(\varphi_0)\varphi_1^2-{1\over 2}{V^\prime(\varphi_0)\over\varphi_0}\varphi_2^2
\eeq
where $a,b=1,2$.
Arbitrary variations $\varphi_1$ and $\varphi_2$ are permitted from the point of view of charge conservation.  All spatial Fourier components in the orthogonal directions will correspond to displacement of charge along the wall in a periodic fashion.  Charge conservation constraints will only apply to the zero frequency Fourier component, that is constant along the wall, and here the variations will have to be constrained.  But this component corresponds to fluctuations that are effectively 1 dimensional, depending only on the coordinate normal to the {\q}-wall configuration.  We know that the {\q}-wall is by definition stable under such fluctuations, hence we will not be concerned with these fluctuations.  
The equations of motion for $\varphi_a$ are
\bar
\ddot\varphi_1-2\omega\dot\varphi_2-\omega^2\varphi_1-\nabla^2 \varphi_1 +V^{\prime\prime}(\varphi_0)\varphi_1&=&0\cr
\ddot\varphi_2+2\omega\dot\varphi_1-\omega^2\varphi_2-\nabla^2 \varphi_2 +{V^{\prime}(\varphi_0)\over\varphi_0}\varphi_2&=&0.
\ear
Replacing $\varphi_a(\vec x,t)\rightarrow e^{-i(\Omega t+k_1 y+k_2z) }
\varphi_a(x)$ and writing the resulting equations in matrix form yields
\beq
\pmatrix{-{d^2\over dx^2} +V^{\prime\prime}(\varphi_0)-\omega^2&i2\omega\Omega\cr -i2\omega\Omega &-{d^2\over dx^2} +{V^{\prime}(\varphi_0)\over\varphi_0}-\omega^2} \pmatrix{\varphi_1\cr\varphi_2}=(\Omega^2 -|\vec k|^2)\pmatrix{\varphi_1\cr\varphi_2}
\eeq
which is a transcendental equation for $\Omega$.

The operator on the LHS plays the role of a Hamiltonian whose eigenvalues determine if a mode is oscillatory or exponential.  The Hamiltonian is a hermitean operator for real $\Omega$, hence its spectrum is real.  It is easy to see that all non-negative eigenvalues of the Hamiltonian at $\Omega =0$ will give rise to oscillatory modes of the fluctuation spectrum.  Only negative modes at $\Omega =0$, for small enough values of $|\vec k|$ can yield exponential modes, but even these become oscillatory modes for large enough $|\vec k|$.  The non-standard aspect of this stability problem is that the  Hamiltonian also directly depends upon the fluctuation frequency $\Omega$.  We must find the spectral curves $E(\Omega )$ as functions of $\Omega$ and then look for solutions of the equation $E(\Omega ) = \Omega^2-|\vec k|^2$.  The RHS is an upward opening parabola for real $\Omega$ with intercept at $-|\vec k|^2$.  The spectral curve of any eigenvalue of the Hamiltonian that is non-negative at $\Omega =0$ will intersect with the parabola at two distinct points.  This is because for large enough $\Omega$ the spectral curves are simply asymptotic to the linear functions $E(\Omega )=\pm 2\omega\Omega $. It is only the eigenvalues of the Hamiltonian  that are negative at $\Omega=0$ which can yield exponentially decaying solutions.  We will see that there is exactly one such mode for this Hamiltonian with the choice of the scalar potential taken in \ref{potential}.
\subsection{Spectrum of the Hamiltonian and supersymmetric quantum mechanics}
The Hamiltonian for $\Omega =0$ is given by
\beq
{\cal H}=\pmatrix{{\cal H}_+ &0\cr 0&{\cal H}_-}=\pmatrix{-{d^2\over dx^2} +V^{\prime\prime}(\varphi_0)-\omega^2&0\cr 0 &-{d^2\over dx^2} +{V^{\prime}(\varphi_0)\over\varphi_0}-\omega^2} .
\eeq
There is an obvious exact zero energy level, 
\beq
\psi^-=\pmatrix{\varphi_1\cr\varphi_2}=\pmatrix{ 0\cr\varphi_0(x)}
\eeq
since $\varphi_0$ satisfies the equation \ref{varphi0}.  Remarkably the mode
\beq
\psi^+=\pmatrix{\varphi_1\cr\varphi_2}=\pmatrix{ \varphi_0^\prime (x)\cr 0}\label{zeromode}
\eeq
is also an exact zero mode, essentially the corresponding eigenvalue equation is the derivative of the equation of motion for $\varphi_0(x)$ given by equation \ref{varphi0}.  The notation will become clear below.  This is, of course, not just a coincidence.  The two sub-Hamiltonians which comprise ${\cal H}$ are actually supersymmetric partners and their spectra are paired except for the lowest energy mode.  Indeed since \ref{zeromode} is the derivative of $\varphi_0(x)$ it has a node, hence it cannot be the lowest energy eigenvalue of ${\cal H}_+$, and there has to be a negative energy mode.
\subsubsection{Stability equations and supersymmetric quantum mechanics\cite{gj}}\label{431}
Consider the class of supersymmetric quantum mechanics models determined by the super potential
\beq
{\cal W}=n\omega\tanh (\omega x).
\eeq
Then the corresponding supersymmetric partner Hamiltonians are given by
\beq
{\cal H}_\pm ={\cal W}^2\mp {\cal W}^\prime =n^2\omega^2-n(n\pm n)\omega^2{\rm sech}^2\omega x
\eeq
The two Hamiltonians ${\cal H}_\pm$ are given in terms of
\beq
{\cal H}_+ = A^\dagger A\quad\quad {\cal H}_-=AA^\dagger .
\eeq
where 
\beq
A={d\over dx} +{\cal W}\quad\quad A^\dagger =-{d\over dx} +{\cal W}
\eeq
which gives 
\bar
{\cal H}_+ &=&-{d^2\over dx^2} +n^2\omega^2-n(n+1)\omega^2{\rm sech}^2\omega x\cr
{\cal H}_- &=&-{d^2\over dx^2} +n^2\omega^2-n(n-1)\omega^2{\rm sech}^2\omega x.
\ear
The ground state is the wave function annihilated by $A$
\beq
A\varphi_{-1}=0
\eeq
which integrates easily to 
\beq
\varphi_{-1}(x)=\varphi_{-1}(0){\rm sech}^n\omega x.
\eeq
Then, if ${\cal H}_+\psi_+ =E_+\psi_+$ then ${\cal H}_-(A\psi_+) =E_+(A\psi_+)$ which shows that all energy levels are (supersymmetrically) paired except for the level annihilated by $A$.  Equivalently if ${\cal H}_-\psi_-= E_-\psi_-$ the ${\cal H}_+(A^\dagger\psi_-) =E_-(A^\dagger\psi_-)$.  This pairing is equivalent to the former except that there is no state annihilated by $A^\dagger$. To find all the levels of ${\cal H}_\pm$ we proceed by changing the value of $n\rightarrow n-1$. This process continues until we reach $n=1$  and the Hamiltonian ${\cal H}_-|_{n=1}=-(d^2/dx^2)+\omega^2$ for which we know the spectrum to be free, plane waves.  The ground states of all intermediate Hamiltonians are given by 
\beq
\psi\sim{\rm sech}^m\omega x\quad\quad m=1,2,\cdots ,n.
\eeq
Then applying the appropriate strings of $A^\dagger_m$ (using an obvious notation) generates all of the bound states of ${\cal H}_\pm $ as shown in detail in \cite{cks}. 
The interesting point is that we can easily find the potential $V(\varphi_0)$ that is necessary to generate this system of supersymmetric Hamiltonians for the stability problem of the {\q}-wall.  Indeed, we impose that
\bar
{V^\prime (\varphi_0)\over\varphi_0}&=&\omega^2n(n-1){\rm sech}^2\omega x +B\cr\label{consistency1}
V^{\prime\prime}(\varphi_0)&=& \omega^2n(n+1){\rm sech}^2\omega x +C\label{consistency2}
\ear
where $B,C$ are constants to be adjusted.  This yields
\beq
{1\over n-1}\big( {V^\prime (\varphi_0)\over\varphi_0} -B\big)= \omega^2n {\rm sech}^2\omega x ={1\over n+1}\big( V^{\prime\prime}(\varphi_0) -C\big) .
\eeq
This equation is equivalent to
\beq
-\big({n+1\over n-1}\big)\varphi_0^{-\big( {n+1\over n-1}\big)-1}V^\prime (\varphi_0) +\varphi_0^{-\big( {n+1\over n-1}\big)}V^{\prime\prime}(\varphi_0) +\big( { B(n+1)-C(n-1)\over n-1}\big) \varphi_0^{-\big( {n+1\over n-1}\big)}=0
\eeq
after multiplying by an integrating factor.  Then this equation is easily integrated to
\beq
V(\varphi_0) = {D\varphi_0^{\big( {n+1\over n-1}\big)+1}\over {\big( {n+1\over n-1}\big)+1}}+\big( { B(n+1)-C(n-1)\over 4}\big) \varphi_0^2 +G
\eeq
for some integration constants $D,G$.  Now imposing consistency with equations \ref{consistency1} yields
\beq
{V^\prime (\varphi_0)\over\varphi_0}=D\varphi_0^{\big( {2\over n-1}\big)}+\big( { B(n+1)-C(n-1)\over 2}\big)=\omega^2n(n-1){\rm sech}^2\omega x +B.
\eeq
Taking $D=\omega^2 n(n-1)$ yields
\beq
\varphi_0={\rm sech}^{n-1}\omega x \quad {\rm and}\quad{ B(n+1)-C(n-1)\over 2}=B
\eeq
On the other hand from \ref{consistency2}
\beq
V^{\prime\prime}(\varphi_0)=D\big( {n+1\over n-1}\big) \varphi_0^{-\big( {n+1\over n-1}\big)-1}+\big( { B(n+1)-C(n-1)\over 2}\big) =\omega^2n(n+1){\rm sech}^2\omega x +C
\eeq
hence taking $D=\omega^2 n(n-1)$ as before is consistent since
\beq
D\big( {n+1\over n-1}\big) =\omega^2 n(n+1)
\eeq
as required.  The constants however must satisfy (along with the equation from before)
\bar
{ B(n+1)-C(n-1)\over 2} &=&B\cr
{ B(n+1)-C(n-1)\over 2}&=&C.
\ear
These are two homogeneous linear equations for $B,C$ which only have the trivial solution except if the system is dependent.  It is indeed, and we find the solution is simply $B=C$.  We see that the potential is a polynomial for only the cases $n=2,3$.  
\subsection{Stability analysis for analytical {\q}-walls}
The system studied in section \ref{aqw} uses
\beq
V^\prime (\varphi_0(x) )=\big( -6\omega^2 {\varphi_0(x)\over\phi_0}+5\omega^2\big)\varphi_0(x)
\eeq
hence replacing $\varphi_0(x)=\phi_0{\rm sech}^2\omega x$ yields
\beq
{V^\prime (\varphi_0(x) )\over \varphi_0(x)}=\big( 5\omega^2-6\omega^2 {\varphi_0(x)\over\phi_0}\big)= 5\omega^2-6\omega^2{\rm sech}^2\omega x
\eeq
while
\beq
V^{\prime\prime}(\varphi_0(x) )=\big( 5\omega^2-12\omega^2 {\varphi_0(x)\over\phi_0}\big)=5\omega^2-12\omega^2{\rm sech}^2\omega x.
\eeq
Thus the system corresponds to the case $n=3$ above with $B=C=5\omega^2$.  The corresponding stability equation is 
\bar
\pmatrix{-{d^2\over dx^2} +4\omega^2-12\omega^2{\rm sech}^2\omega x&i2\omega\Omega\cr -i2\omega\Omega &-{d^2\over dx^2} +4\omega^2-6\omega^2{\rm sech}^2\omega x} \pmatrix{\varphi_1\cr\varphi_2}&=&\cr
(\Omega^2 -|\vec k|^2)\pmatrix{\varphi_1\cr\varphi_2}&&.\label{eigenvalue}
\ear
The spectrum of this Hamiltonian at $\Omega=0$ corresponds to a single negative energy bound state with $E=-5\omega^2$
\beq
\psi_{-5\omega^2}^+=\sqrt{15\omega\over 16} \pmatrix{{\rm sech}^3\omega x \cr 0}
\eeq
two zero energy bound states with $E=0$
\beq
\psi_{0}^+=\sqrt{15\omega\over 4} \pmatrix{{\rm sech}^2\omega x\tanh\omega x \cr 0}\quad\psi_{0}^-=\sqrt{3\omega\over 4} \pmatrix{0\cr {\rm sech}^2\omega x}
\eeq
and two positive energy bound states with $E=3\omega^2$
\beq
\psi_{3\omega^2}^+=\sqrt{24\omega} \pmatrix{12{\rm sech}\omega x-15{\rm sech}^3\omega x \cr 0}\quad\psi_{3\omega^2}^-=\sqrt{3\omega\over 2} \pmatrix{0\cr {\rm sech}\omega x\tanh\omega x}
\eeq
and a continuum starting at $E> 4\omega^2$.  All these modes are found in a straightforward manner using the general ideas schematized in section \ref{431}, for more details see \cite{cks}.  

The eigenvalue problem \ref{eigenvalue} has solutions that are best expressed graphically.  The spectral curve $E(\Omega)$ of each eigenvalue of the Hamiltonian on the LHS of \ref{eigenvalue} must intersect the parabola $\Omega^2 -|\vec k|^2$.  The eigenvalue problem for this Hamiltonian can be written algebraically as
\beq
(H_0+H_1\sigma_3+2\omega\Omega\sigma_2 )\psi =E(\Omega )\psi
\eeq
where $\sigma_i$ are the usual Pauli matrices.  The spectrum is doubly
degenerate for each non-negative mode at $\Omega=0$ due to the
underlying supersymmetry.  Clearly
\beq
(H_0+H_1\sigma_3-2\omega\Omega\sigma_2 )\sigma_3\psi =E(\Omega )\sigma_3\psi
\eeq
hence the spectrum is even in $\Omega$.  However we can without loss
of generality restrict $\Omega >0$ because $\varphi_i$ are in
principle real: reversing the sign of $\Omega\rightarrow -\Omega $ does not generate an independent set of solutions.  

We see clearly that the non-negative modes satisfy $E(\Omega )> \Omega^2 -|\vec k|^2$ for $\Omega\approx 0$, but as $\Omega\rightarrow\infty$ these spectral curves asymptote to $E(\Omega )\rightarrow \pm 2\omega\Omega$, the spectrum of the operator
\beq
2\omega\Omega\sigma_2.
\eeq
These are linear functions of $\Omega$ which for sufficiently large $\Omega$ always satisfy $\Omega^2 -|\vec k|^2>2\omega\Omega$.  Hence by continuity each spectral curve must intersect at least once with the parabola, giving rise to a real solution for $\Omega$, and hence an oscillatory mode in the fluctuation spectrum.  We speculate that this occurs exactly once for the spectral curve of each non-negative eigenvalue.

\begin{figure}[ht]
\epsfysize=6cm
\epsfxsize=6cm
\centerline{\epsfbox{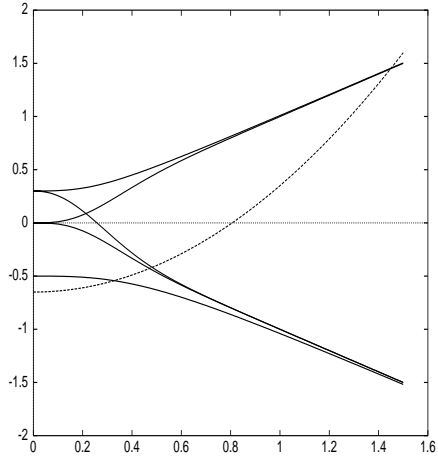}}
\caption{Graphical depiction of transcendental equation for
  $\Omega$ (Eq. \ref{eigenvalue},with an appropriate numerical choice for $\omega$); $k^2>.5$. Solid line represents discrete
  eigenvalues of the LHS of this equation; continuous eigenvalues are
  expected to have similar behavior.  Broken line represents the
  RHS. Points of intersection give the (real) values of $\Omega$ for
  the corresponding modes; no intersection would indicate no real
  solution, indicating an unstable mode of the $Q$-wall. No such
  instabilities exist in this case.}
\end{figure}

\begin{figure}[ht]
\epsfysize=6cm
\epsfxsize=6cm
\centerline{\epsfbox{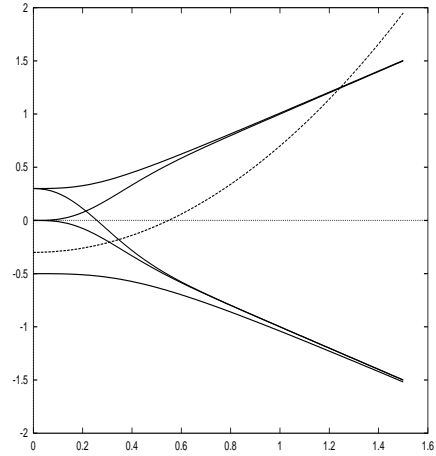}}
\caption{Graphical depiction of transcendental
  equation for $\Omega$ (Eq. \ref{eigenvalue},with an appropriate numerical choice for $\omega$); $k^2<.5$. No solution exists for lowest eigenvalue of the LHS, indicating an unstable mode.}
\end{figure}

Only the negative mode needs to be treated differently.  Here, at $\Omega=0$ and for $|\vec k|^2$ small enough, the negative mode is already below the parabola and hence there is generically no real solution for $\Omega$.  We cannot compute the solution analytically, but a perturbative approach is fruitful.  We can compute the spectral curve to second order in perturbation theory (the first order change vanishes).  We get
\bar
E(\Omega )&=& -5\omega^2 -4\Omega^2\omega^2\sum_{n=1}^\infty{|<\psi_{-5\omega^2}^+|\sigma_2|\psi_n>|^2\over E_n+5\omega^2} + \cdots\cr
&=&-5\omega^2 -\Omega^2 B^2 +o(\Omega^4 )
\ear
where $B^2$ is a calculable positive real number that is independent of $\omega$.  It is not illuminating to compute $B^2$, it is a number of order 1.  Then the equation to be satisfied is
\beq
-5\omega^2-\Omega^2 B^2 =\Omega^2 -|\vec k|^2
\eeq
that is
\beq
\Omega^2  ={-5 \omega^2  +|\vec k|^2\over 1+B^2}.
\eeq
For $|\vec k|^2<5 \omega^2$ this gives $\Omega^2 <0$, {\it i.e.,} the solution for $\Omega $ is imaginary and the corresponding mode is unstable.  Indeed
\beq
\Omega = i\omega\sqrt{5  -(|\vec k|^2/\omega^2)\over 1+B^2}
\eeq
hence as $|\vec k|^2\rightarrow 5 \omega^2$ from above, we get $\Omega\rightarrow 0$.  The lifetime is defined as $T=1/|\Omega | $ hence for small values of $|\vec k|$ we get $T\sim (1/\omega )$.  But as $|\vec k|^2\rightarrow 5 \omega^2$ we get $T\rightarrow\infty$.  Thus the very long wavelength spatial Fourier modes decay with lifetime of the order of $1/\omega$ while as the wavelength $1/|\vec k|$ approaches $1/5\omega$, which is of the order of the thickness of the \q-wall, the lifetime becomes arbitrarily long, the modes become quasi stable.  For larger $|\vec k|$ (shorter wavelength) the modes are of course strictly stable.  Analysis of related, exactly solvable simple models such as $H(\Omega )\psi=(k^2 +\alpha\sigma_3+\Omega\sigma_2)\psi=\Omega^2 \psi$ yields the same kind of behavior, generically.  
Hence we can conclude that the fluctuation spectrum corresponds to stable oscillatory modes except for one degree of freedom.  The corresponding mode is unstable only if the wavelength of the spatial Fourier mode is larger than the width of the \q-wall.  Furthermore, the lifetime of the unstable mode is larger than $1/\omega\sim Q_1^{(1/2)}=(Q/L^2)^{(1/2)}$, which clearly can be taken to be arbitrarily large.  Hence we conclude that \q-wall, or equally well a \q-string can be arbitrarily long-lived.
\section{Discussion and conclusion}
In this paper we have studied lower-dimensional \q-balls
and extended objects  based on them
in models with a scalar field potential
with a flat direction for large field values.
The energy of a \q-ball of charge $Q_d$ in $d$ space dimensions
varies as $Q_d^{(d/d+1)}$. Thus, as three-dimensional \q-balls
become more and more energetically advantageous with increasing
charge, the same is true (only more so) in two or in one dimension.

The extended configurations, called \q-strings and \q-walls, 
also have interesting energetics. A string of fixed
charge and of varying length $L$, for example,
has energy $E\sim L^{1/3}$, indicating a rapidly decreasing
tension as the length is increased. The time scale of oscillations
turns out to be of the same order as that for a topological string.

These strings and walls have a different instability not shared by
topological strings: namely, density fluctuations along the string or
wall can form, causing a sort of beading: the extended object would
evolve into a one- or two-dimensional array of smaller \q-balls.

This instability was examined in detail for \q-walls, where
supersymmetric quantum mechanics enables a rather detailed
analysis. Only modes corresponding to long-wavelength fluctuations 
are unstable, and the time scale of growth of these fluctuations can
be rather large (compared to the characteristic time scale of the
\q-wall).

It is not clear whether or why such objects would form in the early
Universe. (Unlike conventional cosmic strings, there is no topological 
impetus for their formation.) But if they did form, they could be
long-lived, and could help contribute to structure formation in the
Universe. Such a possibility merits further investigation.

\acknowledgments
We thank NSERC of Canada for financial support.

\end{document}